\documentclass[10pt]{article}
\pdfoutput=1
\usepackage{epsf}

\usepackage{epsfig,graphics}
\usepackage {graphicx}
\usepackage {epsfig}
\usepackage {subfigure}
\usepackage {tabularx} 
\usepackage{rotate}	
\usepackage{slashed}

\usepackage{amsmath}
\usepackage{amsfonts}
\usepackage{amssymb}
\usepackage{graphicx}
\usepackage{cite}
\usepackage{url}
\usepackage{hyperref}
\hypersetup{colorlinks=false,linkcolor=black}
\usepackage{pgfplots}
\pgfplotsset{compat=newest}

\definecolor{myblue}{cmyk}{0.65, 0.37, 0.0, 0.19}
\newcommand{\bmat}{\left(\begin{array}}
\newcommand{\emat}{\end{array}\right)}

\def\yzero{\smash{\hbox{$y\kern-4pt\raise1pt\hbox{${}^\circ$}$}}}

\def\beq{\begin{equation}}
\def\eeq{\end{equation}}
\def\beqa{\begin{eqnarray}}
\def\eeqa{\end{eqnarray}}

\def\-{\hphantom{-}}
\def\ov{\overline}
\def\s2{\frac{1}{\sqrt2}}

\def\beq{\begin{equation}}
\def\eeq{\end{equation}}
\def\beqa{\begin{eqnarray}}
\def\eeqa{\end{eqnarray}}

\def\IF{\relax{\rm I\kern-.18em F}}
\def\II{\relax{\rm I\kern-.18em I}}
\def\IP{\relax{\rm I\kern-.18em P}}
\def\IC{\relax\hbox{\kern.25em$\inbar\kern-.3em{\rm C}$}}
\def\IR{\relax{\rm I\kern-.18em R}}

\def\Dsl{\,\raise.15ex\hbox{/}\mkern-13.5mu D} %this one can be subscripted
\def\IZ{Z\kern-.4em  Z}

\catcode`\@=11   
\newdimen\@rotdimen
\newbox\@rotbox  

\def\@vspec#1{\special{ps:#1}}%  passes #1 verbatim to the output
\def\@rotstart#1{\@vspec{gsave currentpoint currentpoint translate
   #1 neg exch neg exch translate}}% #1 can be any origin-fixing transformation
\def\@rotfinish{\@vspec{currentpoint grestore moveto}}% gets back in synch 
\def\@rotr#1{\@rotdimen=\ht#1\advance\@rotdimen by\dp#1%
   \hbox to\@rotdimen{\hskip\ht#1\vbox to\wd#1{\@rotstart{90 rotate}%
   \box#1\vss}\hss}\@rotfinish}
\def\@rotl#1{\@rotdimen=\ht#1\advance\@rotdimen by\dp#1%
   \hbox to\@rotdimen{\vbox to\wd#1{\vskip\wd#1\@rotstart{270 rotate}%
   \box#1\vss}\hss}\@rotfinish}%
\def\@rotu#1{\@rotdimen=\ht#1\advance\@rotdimen by\dp#1%
   \hbox to\wd#1{\hskip\wd#1\vbox to\@rotdimen{\vskip\@rotdimen
   \@rotstart{-1 dup scale}\box#1\vss}\hss}\@rotfinish}%
\def\@rotf#1{\hbox to\wd#1{\hskip\wd#1\@rotstart{-1 1 scale}%
   \box#1\hss}\@rotfinish}%
\def\rotate{\@ifnextchar[{\@rotate}{\@rotate[l]}}
\def\@rotate[#1]#2{\setbox\@rotbox=\hbox{#2}\@nameuse{@rot#1}\@rotbox}

\catcode`\@=12
\topmargin
-1.5cm
\textwidth
15.5cm
\textheight
23.5cm
\oddsidemargin
0.7cm
\evensidemargin
0.7cm

\begin{document}

%----------------------------------------------------------------------%
%  numbering equations with section number
%----------------------------------------------------------------------%
\makeatletter
\@addtoreset{equation}{section}
\makeatother
\renewcommand{\theequation}{\thesection.\arabic{equation}}
%----------------------------------------------------------------------%
%  title page
%----------------------------------------------------------------------%
\pagestyle{empty}
%\vspace*{1.0in}
\vspace{-0.2cm}
\rightline{IFT-UAM/CSIC-18-039}
\rightline{April 19-th 2018}
%\rightline{\tt hep-th/xxxxxxx}
\vspace{1.2cm}
\begin{center}
%\vspace{0.5cm}
\LARGE{  A Note on 4D  Heterotic String Vacua, \\
FI-terms and the Swampland} 
%({\it Questo e una genialit\'a o e tutto una cassata})
\\[13mm]
  \large{    Gerardo Aldazabal$^1$ and Luis E. Ib\'a\~nez$^2$,
   \\[6mm]}
\small{
$^1$  {\em G. F\'isica CAB-CNEA, CONICET and Instituto Balseiro 
}\\{\em Centro At\'omico 
Bariloche, Av. Bustillo 9500, Bariloche, 
Argentina.}
  \\[0pt]
  $^2$ {\em Departamento de F\'{\i}sica Te\'orica
and Instituto de F\'{\i}sica Te\'orica UAM/CSIC,\\[-0.3em]
Universidad Aut\'onoma de Madrid,
Cantoblanco, 28049 Madrid, Spain} \\[0pt]
}
\small{\bf Abstract} \\[7mm]
\end{center}
\begin{center}
\begin{minipage}[h]{15.22cm}
We present a conjecture for the massless sector of perturbative  4D $N=1$ 
heterotic $(0,2)$ string vacua, including $U(1)^n$ gauge symmetries,
one of them possibly anomalous (like in  standard heterotic compactifications).
Mathematically it states that the positive  hull generated by the charges of the 
 massless chiral multiplets spans a sublattice of the 
full  charge lattice. We have tested this conjecture in many heterotic $N=1$ 
compactifications  in 4D.
Our motivation for this conjecture is that it allows to understand a very old 
puzzle  in $(0,2)$ $N=1$ heterotic compactification
with an anomalous $U(1)$.  The conjecture guarantees that 
 there is always a D-flat direction cancelling the FI-term 
and restoring   $N=1$ SUSY in a nearby vacuum.  This is something that has 
being 
verified in the past in a large number of cases, but 
whose origin has remained obscure for decades.
We argue that the existence of this lattice of massless  states guarantees the 
instability of heterotic non-BPS  extremal
blackholes, as required by Weak Gravity Conjecture arguments. Thus the pervasive 
existence of these nearby FI-cancelling vacua 
would be connected with WGC arguments.

\end{minipage}
\end{center}
\newpage
%----------------------------------------------------------------------%
%  Resetting of counters
%----------------------------------------------------------------------%
\setcounter{page}{1}
\pagestyle{plain}
\renewcommand{\thefootnote}{\arabic{footnote}}
\setcounter{footnote}{0}
%----------------------------------------------------------------------%
%  Paper begins
%----------------------------------------------------------------------%

%&&&&&&&&&&&&&&&&&&&&&&&&&&&&&&&&&
\section{Introduction}
%&&&&&&&&&&&&&&&&&&&&&&&&&&&&&&&&&
%\newpage

Four dimensional string vacua often have  a number of gauged $U(1)$ symmetries. 
Some of them are sometimes anomalous 
with anomalies cancelled by the 4D version of the GS mechanism.  In heterotic 
vacua obtained from CY with non-Abelian bundles
or standard  $(0,2)$ Abelian orbifolds  at most one anomalous 
$U(1)_X$ is allowed, whose anomaly is cancelled by the shift transformation 
of
the axionic partner of the axi-dilaton  Im$S$.   Supersymmetry then tells us 
that there is an associated FI-coupling\cite{FIold} such that one has 
D-term of the form
\beq
V_X \ =\ \frac {1}{S+S^*}\left[ \xi _X\ +\ \sum_i q_{X,i}|\Phi_i|^2\right]^2 \ ,
\label{pot}
\eeq 
with
\beq
\xi _X\ =\ \frac {TrQ_X}{48(2\pi)^2 \kappa_4^2} \frac {1}{S+S^*} \ .
\eeq
Here the sum runs over all scalars in the theory charged under the anomalous 
$U(1)$, and $TrQ_X$ is the trace over all massless charged 
chiral multiplets in the theory.
Perturbatively the potential of the dilaton is flat and,  for non-vanishing 
values a field-dependent FI-term, $\xi_X\not=0$ appears to break 
SUSY. From the very early days of heterotic 
compactification  it was found  that,  in any such $(0,2)$ 4D 
heterotic vacua, there is always a nearby SUSY vacuum
 in which some scalars $\Phi_i$ with  the correct charge get appropriate 
vevs to get a vanishing D-term in a new SUSY vacuum.   However, the reason why 
this is true was never fully understood in the literature.
In the present paper we come back to this puzzle and take advantage of recent 
efforts\cite{WGC1,WGC2,WGC3,WGC4}  to sort out the set of theories which may be
 embedded into a consistent theory of quantum gravity from those which cannot 
and belong to the {\it swampland}\cite{swampland,WGC,WGCM} (see 
\cite{vafafederico} for a review).
 We  argue that putative theories in which the FI-term is not 
cancelled would inconsistent or belong to the {\it swampland}.

It is well known that a description in terms of holomorphic 
scalar operators\cite{Buccella} provides a useful way to look for D-flat 
directions for an arbitrary gauge group in $N=1$ SUSY.
Having  a flat direction cancelling all the the D-terms  corresponds to the 
existence of an operator involving the scalar components
of massless chiral scalar  fields $\phi_i$ (but not the conjugates)
\beq
I\ =\ (\phi_i\phi_j...\phi_k) 
\eeq
such that is has  $U(1)_X$ charge opposite to $TrQ_X$ and zero charge with 
respect to any other charge or gauge interaction. 
Setting the vevs fulfilling 
\beq
\frac {\partial I}{\partial \phi_i}\ =\ c\phi_i^* \ ,
\label{cancel}
\eeq
all D-terms vanish (c is a constant). Note that the operators $I$ are just a 
book-keeping device to see what fields may have simultaneously 
vevs  with a vanishing D-term, they do not need to be present in the effective 
action.
 What experience tells us is that 
such operators with the required sign of the $Q_X$ always exist in any $(0,2)$ 
heterotic model so far analyzed.
This suggests that there could be some deeper principle why this is always the 
case. It could  well be, in particular,  that 
consistency with quantum gravity forces this to happen, for some reason still 
to be understood. 

It is reasonable to ask whether the existence of scalars with the correct 
charge, opposite to the FI-term, is a direct consequence of
anomaly cancellation and $N=1$ SUSY. After all,  the cancellation of the mixed 
$U(1)$-gravitational anomalies seems to require the presence of
fermions with opposite charges. However this is not so when the $U(1)$ is 
anomalous. In the class of theories under
consideration the coefficients of the  cube and the mixed-gravitation  
anomalies of $U(1)_X$ are related by Green-Schwarz anomaly
cancellation constraints (see e.g.\cite{BOOK} and references therein)
\beq
\frac {1}{3k_X}\sum_iq_{X,i}^3 \ =\  \frac {1}{24} TrQ_X  \ ,
\label{anom}
\eeq
where $k_X$ is the normalization of the $U(1)_X$ coupling.
In principle this  may be fulfilled with all scalars having the same sign, so 
that the FI-term would never cancel. For example one
may have a model with two chiral multiplets with charge  $q_1=1, q_2=1/2$. The 
reader may check that for $k_X=6$ 
Eq.\eqref{anom} is fulfilled, and anomalies cancel through the GS mechanism. 
We 
intend to put forth  that this cannot happen in 
$N=1$ heterotic vacua and such model would be in the swampland. It seems that 
anomaly cancellation is not strong enough to
guarantee FI-term cancellation.

In Section \ref{sec:The positive cone conjecture} we formulate a conjecture 
concerning massless sector of 
$N=1$ heterotic vacua and provide some examples. In Section 
\ref{sec:Blackholes, 
the WGC and  FI-terms} we discuss our conjecture  and its possible connection to 
 the  Weak 
Gravity Conjecture (WGC)\cite{WGC}. We conclude with some comments  in Section 
\ref{sec:comments}

\section{The positive cone conjecture}
\label{sec:The positive cone conjecture}

Consider a $N=1$ $D=4$ heterotic vacuum with  gauge group $G=H\times U(1)^N$, 
where $H$ is some semisimple group and where a combination $U(1)_X$ of the  
$N$ $U(1)'s$ may be anomalous. There are 
massless chiral multiplets with complex scalar components
with a vector of charges
\beq
\phi_i \ =(R_i; q_1^i,..,q_N^i) ,
\eeq
where $i$ runs over all the massless chiral spectrum. Here $R_i$ is some 
representation of the non-Abelian semisimple group $H$.
By holomorphically multiplying these scalars it is possible to  obtain 
operators  $\Phi^a$ which are singlets under the non-Abelian group 
\beq
\Phi^ a\ =\ (\phi_i.....\phi_k)^a
\eeq
each one with a vector of charges
\beq
q_a \ =\  (q_1^a,..,q_N^a)  \ .
\eeq
Consider now all the vectors of charges generated by:
\beq
\Lambda_0 \ =\ \{ \sum_a M_a q_a \ ,\ M_a\in {\bf Z^+}\}
\eeq
where $a$ runs over all possible $\Phi^a$ chiral operators. $\Lambda_0$ is the 
positive hull generated by the charges of all massless chiral fields.
The conjecture is then:

{\it The positive hull $\Lambda_0$ generated by the charges of all massless 
chiral fields is a sublattice of the full charge lattice $\Lambda$}.

The important point here is that $\Lambda_0$,  being a sublattice,  contains a 
given vector {\it and} its opposite.
Note that the statement is non-trivial, in principle it could had happened that 
$\Lambda_0$ {\it was not} a lattice, but just a set of vectors,
not including the opposite of each vector. In fact,  the positive 
hull of the example given above with two charged particles is not
a lattice.

If the conjecture was true, there should always be a flat direction at 
which the FI-term would be cancelled.  Indeed,  to each member of the
sublattice $\Lambda_0$  (choice of integers $M_a$) corresponds an operator 
$(\phi_1...\phi_N)$. In particular, $\Lambda_0$ should contain
an operator $I_X$ which is only charged under $U(1)_X$  {\it and} another 
operator $I_{-X}$
with opposite charge.   Either one or the other will be able to cancel the 
FI-term 
by assigning vevs as in eq.(\ref{cancel}). So the existence of these sublattice 
would guarantee the
existence of a D-flat direction cancelling the FI-term and preserving $N=1$.

  We now discuss examples of heterotic 
compactifications, 
showing how  sublattices always arise in the massless sector
of the theory.

The reader uninterested in the details of  these models may safely jump to 
Section \ref{sec:Blackholes, the WGC and  FI-terms}.

\subsection{Examples}

We have tested this conjecture in many Abelian  $(0,2)$ $Z_N$ orbifolds of the 
heterotic $E_8\times E_8$ and $SO(32)$ strings leading to 
chiral $D=4$,$N=1$ theories with or without an anomalous $U(1)$. Here we show a 
couple of representative examples (see e.g.  \cite{BOOK} for a review on 
heterotic orbifold constructions) and 
present further ones in the Appendix.  

\subsubsection{\texorpdfstring{ $Z_3$ orbifold \,}  \texorpdfstring{$E_8\times 
E_8$}  examples }

A  simple example with a single anomalous $U(1)$ is the $Z_3$ orbifold 
with shift $V=1/3(11112000)\times (20000000)$ acting on the  $E_8\times E_8$ 
gauge lattice. 
 This model has gauge group  $SU(9)\times SO(14)\times U(1)_X$ and charged  
massless chiral spectrum  given by 
\begin{eqnarray}
U&:&\ \ 
%\longrightarrow& 
\ 3[  ({\bf 84} ,{\bf 1})_0\ +\ ({\bf 1},{\bf 14} )_{-1}\ + 
\ ({\bf 1},{\bf 64} )_{1/2}]   ;\\ 
T&: &\ \
%&\longrightarrow &\ 
27((\overline {\bf 9},{\bf 1} )_{2/3} ,
\end{eqnarray}
where $U$ and $T$ denote untwisted and twisted spectrum and the subindex is the 
charge under the $U(1)$ generator
$Q_X=(1,0,..,0)$ in the second $E_8$. In this simple case the sublattice is 
generated by 
\beq
\Lambda_0 \ =\ \left(  M(\pm 2) \ ,\ M\in {\bf Z}\right)  \ .
\eeq
The minimum charge for this lattice comes from operators like 
$({\bf 1},{\bf 14} )^2_{-2}$ , $({\bf 1},{\bf 64} )^4_2$ , $[(\overline {\bf 
9},{\bf 1} )^3 ({\bf 84} ,{\bf 1})]_2$, etc.
Here $TrQ_X = 24\times 9$ and the FI could be cancelled with 
vevs
corresponding to the operator $(14)^2_{-2}$, and $SO(14)$  is broken to 
$SO(12)$.

A simple model with two $U(1)$'s is provided by the $Z_3$ orbifold  with 
embedding 
$V=1/3(110..0)\times (200..0)$ yielding gauge group 
$E_7\times U(1)_X\times SO(14)\times U(1)$. The chiral spectrum is given by
\begin{eqnarray}
U&:&\ \ 
%U&\longrightarrow&
3[({\bf 56} ,{\bf 1})_{1,0}\ +\ ({\bf 1} ,{\bf 
1})_{-2,0}] \ +\ 3[ ({\bf 1} ,{\bf 14})_{0,-1}\ +\ ({\bf 1},{\bf 64} 
)_{0,1/2} ]\\
T&:&\ \ 
%T&\longrightarrow& 
27[  ({\bf 1},{\bf 14} )_{2/3,-1/3} \ +\  ({\bf 
1},{\bf 1})_{2/3,2/3} \ +\ ({\bf 1},{\bf 
1})_{-4/3,2/3} ] \ ,
\end{eqnarray}
and the first $U(1)_X$ is anomalous.
A  sublattice is given by:
\beq
\Lambda_0 \ =\  \{  M\times (4/3,-2/3) \ +\ N\times (2/3,2/3) \ ,\ M,N\in {\bf 
Z} \} \ .
\eeq
In this case it is generated by single twisted fields but there is a smaller 
lattice generated from
the untwisted fields with vector charges 
%and generated by 
$(\pm2,0),(0,\pm2)$.  
coming  from the operators
$[({\bf 56} ,{\bf 1})]^2_{2,0}$,$({\bf 1} ,{\bf 
1})_{-2,0}$ and  $[({\bf 1},{\bf 64})]^4_{0,2} ]$, 
$[({\bf 1} ,{\bf 14})]^2_{0,-2}$ respectively.
In this example $TrQ_X=18\times 24$ so that the FI may be simply cancelled by 
the 
untwisted singlet $({\bf 1} ,{\bf 
1})_{-2,0}$ which is already in the massless sector. However cancellation 
can be also achieved by e.g. the
operator $[({\bf 1} ,{\bf 14})_{0,-1}^2 ({\bf 1},{\bf 
1})_{-4/3,2/3}^3]_{-2,0}^3$ which 
involves twisted states. The normalizations are $k_X=4$, $k'=2$ from 
$Q_X=(1,1,0\dots0)$, $Q'=(1,0,0\dots0)$.

In the above examples the FI-term could be cancelled by using operators/fields 
making use only of  untwisted fields.
However there are plenty of examples in which  the untwisted subsector does 
not generate a sublattice  by itself,
and a full sublattice only arises from the complete untwisted and twisted 
spectrum. The following example has this property.

\subsubsection{An   \texorpdfstring{$SO(32), \, Z_7$ example}{} }
Consider $v=\frac17(1,2,-3)$ and a gauge shift with embedding $ 
V=\frac17(3,\dots,3,0,0)$
leading to gauge group $SU(14)\times U(1)\times SO(4)$ with $Q_X=(1,1,\dots 
1;0,0)$. The untwisted sector massless field content is generated by 
left handed lattice momenta (underlining means all 
possible permutations) $P= ({\underline {- 1,-1,0,0,\dots};0,0})$ with 
$PV\equiv 
\frac{1}7$ and   $({\underline 
{- 1,0,0,\dots};{\underline {\pm 1,0}}}) $ with $PV\equiv 
-\frac{3}7$ and reads
\begin{eqnarray}
U&:&\ \ (\ov {\bf 91},{\bf 1})_{-2}\nonumber  +({\bf 14},{\ov  
{\bf 4}} )_{(-1)}
 \end{eqnarray}
with no massless states in the $\frac27$ sector.
The $m$ twisted sector ($m=1,2,4$)  massless states can be read from the $P$ 
states satisfying $\frac{(P+mV)}2+E_0+N_L-1=0$ with $E_0=\frac27$ and 
$N_L=0,\frac17,\frac27,\frac37, \frac47$ the left oscillator number with 
 associated multiplicities $1,1,2,3,5$ respectively.
Then one finds the twisted chiral fields
\beq
T:\ \
7[ 3({\bf 
1},{\bf 2})_{(-1)}+({\bf 14},{\bf 
1})_{(0)}+5({\bf 
1},{\bf 1})_{(-2)}+ ({\bf 1},{\bf 
2})_{(3)}+({\bf 
1},{\bf 1})_{(-4)}]
\eeq
Notice that $\frac1{24}TrQ_X=-14= \frac1{28.3}TrQ^3_X$.
The  singlets 
\begin{equation}
[({\bf 1},{\bf 2})_{(3)}^2({\bf 1}, {\bf 
1})_{(-4)}]_{(2)}; ({\bf 
1},{\bf 1})_{(-2)}
\end{equation}
generate the sublattice $\{2,-2\}$.
The singlets  $[({\bf 1},{\bf 2})_{(3)}]_{(6)}^2$ or $[({\bf 1},{\bf 
2})_{(3)}^2({\bf 1}, {\bf 
1})_{(-4)}]_{(2)}$ constructed up from the only positive charge 
massless field $({\bf 1},{\bf 2})_{(3)}$ could be used to cancel 
FI term. In this example only the twisted sector had fields with positive 
charge 
and hence the untwisted fields cannot generate a sublattice by themselves  . 
Further examples are presented in the Appendix.

\section{Blackholes, the WGC and  FI-terms}
\label{sec:Blackholes, the WGC and  FI-terms}

We see that the existence of the above sublattice guarantees  
D-flat directions 
in which the FI-term in the potential is cancelled by the vev of opposite 
charge scalars. We will 
now  argue that the WGC could be at the origin of the existence of this 
sublattice.  We do not have any
formal proof of this statement but we want to present in this section some 
circumstantial evidence in this
direction.

In general terms, the Weak Gravity Conjecture  \cite{WGC,WGC1,WGC2} (see 
\cite{vafafederico} for a review)
states that  gravity is the weakest force. In the context of a $U(1)$ gauge 
theory,
it states that any (non-BPS) extremal charged blackhole should be able to decay 
into a superextremal particle  with
mass $m < Q$ in Planck units.  This is required if we want to avoid a tower of 
remnant stable extremal blackholes which are
problematic from different points of view.  Interestingly,  the toroidal 
compactifications of the
heterotic strings down to $4D$ provide the prototypical example in which indeed 
the appropriate superextremal
particles exist with the appropriate characteristics.   A. Sen first described 
in \cite{Sen:1994eb}  the structure of extremal blackholes 
in heterotic  toroidal compactifications.  There are extremal BPS blackholes 
with masses
$m^2=P_R^2/2$ in Planck units and extremal non-BPS blackholes with masses 
$m^2=P_L^2/2$.
On the other hand the
 spectrum of masses of the heterotic 
string states is given by the expression
\beq
\alpha' M^2 = \alpha' 2M_L^2 = 4 \left( \frac {P_L^2}{2} \  +\ N_L\ -\ 1\right) 
\ =\  4\left( \frac{P_R^2}{2} \ +\ N_R\right) \ \ .
\eeq
Here $P_L$ and $P_R$ are the left- and right-handed momenta. They span lattices 
with signature $(22,6)$, with $P_L$ including
the $E_8\times E_8$ or $Spin(32)$ gauge degrees of freedom. 
For $P_R^2>P_L^2$ one finds BPS states for $N_R=0$ and 
 $N_L=\frac {1}{2}(P_R^2-P_L^2)+1$.  They have mass $M^2=P_R^2/2$.
 In addition, for $P_L^2>P_R^2$ and $N_L=0$, $N_R=(P_L^2-P_R^2)/2-1$ one
 has non-BPS states with mass $M^2=P_L^2/2-1$. 
 The masses of these non-BPS states tends to $M^2=P_L^2/2$ for large charges. 
 This nicely fits with the spectrum of blackholes found in \cite{Sen:1994eb}.
 It also shows an explicit realization of the WGC bounds. Indeed, as we go to 
 smaller values of the charge we find string states obeying the WGC with the 
inequality saturated
 for the BPS states with $P_R^2>P_L^2$.  For $P_R^2<P_L^2$ however the extremal 
blackholes 
 have $m^2=P_L^2/2$ whereas there is always a lighter string state with 
$M^2=P_L^2/2-1$.
 This canonical example of heterotic realization of the WGC was first presented 
in \cite{WGC}.

This example has $N=4$ whereas the theories that we are studying have $N=1$ and 
are chiral. However, in the case of toroidal orbifold compactifications
we might  expect that, at least  in the untwisted sector of the 
theory, towers of non-BPS extremal blackholes with masses
$M^2/4=P_L^2/2$ still remain in the spectrum. 
%However due to the twist,  the 
%spectrum of string states has changed and  in general the
%simple structure with one string state with mass  just below the corresponding 
%blackhole with the same charge does not always survive.
Interestingly, in some cases the wanted  instability requires 
the existence of {\it massless} chiral fields in the spectrum, and these
required massless fields have the correct charge to cancel a FI-term in the 
potential, giving a connection between BH instability and FI-term
cancellation.

An example in which this happens is the simple
$Z_3$ orbifold with gauge group $E_7\times U(1)_X\times SO(14)\times U(1)$ 
discussed above. Consider the  $E_8$ lattice vector $P_L=(-1,-1,0,0,0,0,0,0)$.
Associated to this there would be an extremal non-BPS blackhole with mass
$m^2/4=P_L^2/2$ and charges $(-2,0)$ with respect to the $U(1)'s$.
 But precisely for this lattice vector there is a massless chiral field
with the same charges   ${\bf 1}_{-2,0}$, verifying WGC bounds.
On the other hand, as explained in the previous 
section,
this singlet can cancel the FI-term associated to the anomalous $U(1)$. So this 
is an example in which the massless singlet plays a double roll of insuring 
WGC constraints  and FI-term cancellation.  This is just an example, and 
there are many others.  In many of them  however there are no appropriate 
single field states  in the untwisted massless 
sector that could play this role. In general multi-particle states both from the 
untwisted
and twisted massless field sectors are needed to build  the required sublattice. 
But 
at least these simplest examples show the  possible connection between 
the need for states to verify the WGC and the presence of the required massless 
fields
to cancel the FI term.

Another important point to take into account is the corrections to  mass/charge 
ratio in non-BPS extremal blackholes.
 It has been shown that 
corrections involving 4-derivative interactions drive $M^2<Q^2$ \cite{KMP}.
In fact those authors find that for $D=4$  the 
corrected mass of extremal back holes in the heterotic string is
\beq
\frac {M^2}{M_p^2} \ =\   Q^2 \ -\frac {3h\Omega_2^2}{20} \frac 
{M_p^2}{M_s^2}
\eeq
where $\Omega_2=2\pi^{3/2}/\Gamma(3/2)$ and $h$ is the dilaton. 
The correction is always negative and may become   numerically important as 
the compact volume increases. Analogous corrections are expected to arise for 
the case here considered of $N=1$ compactifications.
If this was the case the risk of extremal blackholes 
becoming lighter than their prospective string states into
which they could decay appears,  rendering them stable.  A cure to this 
possible 
disease would be that the massless chiral sector of the theory is  
sufficiently rich so that all extremal blackholes can decay always at least to 
sets of massless  (typically  multiplarticle) states. The sublattice structure
of the states spanned by the massless sector would provide for the appropriate 
decay products.
Note that in addition to the towers of extremal blackholes associated to the 
untwisted sector one would also expect
extremal blackholes with
a charge lattice generated by the shifted lattice $(P+n_iV^i)$.  These typically
correspond  to fractional charges.
For these additional blackholes not  to be stable  (typically multiparticle) 
states constructed using twisted massless 
chiral fields would then exist.  Summarizing, we conjecture that for any node 
in the sublattice generated 
by the massless chiral fields  an extremal blackhole with the 
same charge  should exist.  The existence of the
sublattice would then guarantee both extremal BH instability and cancellation 
of the FI-term in D-flat directions.

\section{Comments}
\label{sec:comments}

The above conjecture  for the existence of a lattice $\Lambda_0$  is only a 
{\it sufficient  condition}  for
the FI-term cancellation. In order for the new shifted vacuum to be 
supersymmetric the corresponding 
D-flat direction should also be F-flat. It would be interesting to prove that 
the presence of the appropriate
(multiparticle) decay channels of the blackholes would, in 
addition,  force the
cancellation of F-terms.

The sublattice of states discussed in this note  is reminiscent of  
the sublattice of  $U(1)^N$ charges 
discussed in \cite{WGC1}.   In the third paper in there
it was conjectured that in a theory of quantum gravity with multiple
$U(1)$'s a sublattice of the charge lattice with a 
superextremal particle at every site  must exist. As made clear  in 
\cite{Heidenreich:2017sim},
this can only be true in more than 4D because there are plenty of examples in 
4D  in which only massless particles 
may be superextremal. But then we would have an infinite number of massless 
particles.   In our case this is not what happens.
There are no infinite particles but rather a charge lattice generated by a 
finite number of massless fields. At each node there is 
a multiparticle state to which potential non-BPS extremal blackholes could
decay into.

A natural question is whether the conjecture of the existence of the positive 
cone sublattice should apply to
all $N=1$ string vacua. It seems that the answer is no, and indeed it is easy 
to find e.g.  Type I or Type IIA orientifolds with
$Dp$-branes is which the conjecture does not work. There are a number of 
reasons for this to be the case. Consider for example
the Type I  duals of the $Spin(32)$ heterotic models
\footnote{See as an example the heterotic $Z_3$, $U(4)^4$ model  in 
\cite{finite} and its Type I dual. 
The  massless twisted states in the heterotic side do generate a sublattice. In 
the Type I side the 
massless chiral fields do not.} . Unlike the heterotic 
case, in the  perturbative Type I duals there are no towers of
non-BPS blackholes and there are no spinorial states either.  The duals of the 
$Spin(32)$ lattice and the spinorials appear 
only at the non-perturbative level from the dynamics of D1-branes, which 
decouple in the perturbative regime.  So an argument for a
sublattice based on the stability of extremal blackholes does not hold. 
This is also consistent with the different structure of 
anomalous $U(1)$'s in Type I orbifold vacua. Indeed in the latter class of 
models there can be more than one anomalous $U(1)$
and the  multiple FI-terms associated to those are related to the twisted 
blowing-up modes rather than to the overall dilaton \cite{Ibanez:1998qp}.
These blowing up modes can be put to zero without generating a decoupling of 
the anomalous $U(1)$ couplings whatsoever.  The same is expected to happen in 
Heterotic compactifications with $U(N)$ bundles (see \cite{blumen} and 
references therein).

%\end{document}

\
\section*{Acknowledgments}

We thank   A. Font,  F. Marchesano, M. Montero, A. Sen, A. Uranga  and I. 
Valenzuela  for useful 
discussions. 
This work is partially supported by the grants  FPA2012-32828 from the MINECO, 
the ERC Advanced Grant SPLE under contract ERC-2012-ADG-20120216-320421 and the 
grant SEV-2016-0597 of the ``Centro de Excelencia Severo Ochoa" Programme and 
also by  CONICET grant PIP-11220110100005  and   
PICT-2016-1358.
G. A. thanks the Instituto de F\'isica Te\'orica (IFT
UAM-CSIC) for hospitality and support.

\newpage

\appendix

\section{ \texorpdfstring{$(2,2)$\, compactifications  of }
\texorpdfstring{$SO(32)$} heterotic on a CY }

In any such compactification the gauge group is generically $SO(26)\times 
U(1)_X$, with a massless spectrum given by
\beq
b_{11} ({\bf 26}_1\ + \ {\bf 1 }_{-2}) \ +\ b_{12}({\bf 26}_{-1}\ +\ {\bf 1 
}_2) \ ,
\eeq
where the subindex is the $U(1)_X$ charge in some integer normalization.
We see that $TrQ_X=24(b_{11}-b_{12})$, which is in general non vanishing.
But note that for any values of the Betti numbers, the sublattice generated is 
generated by
\beq
\Lambda_0 \ =	 (2,-2) \ .
\eeq
In particular e.g. even if $b_{12}=0$ , $\Lambda_0$ contains not only $(-2)$, 
but also $(+2)$ from the  $SO(26)$ singlet operator
$({\bf 26}_1)^2$.  In this case the sublattice has index 2 with respect to 
the full charge lattice $\Lambda$ generated by $(\pm1)$.
In fact the index is larger since in the massive spectrum there will be 
spinorial states which will have seminteger charges.
This case is relatively trivial since there is always the required singlet 
scalar ${\bf 1 
}_{\pm2}$ already in the massless sector of the theory,
one does not need several scalars to cancel the FI. This is in general not the 
case, as the following examples show.

\section{\texorpdfstring{$(2,2)$ compactification of }
\texorpdfstring{$SO(32)$} heterotic on  \texorpdfstring{$Z_3$  orbifold}{} }

In this case there is a gauge shift embedding:
\beq
V\ =\ \frac {1}{3}(11200...0) \ ,
\eeq
and the gauge group is $SO(26)\times SU(3)\times U(1)_X$.
The chiral spectrum   contains from untwisted and twisted sectors:
\begin{eqnarray}
U&:&\ \ 
%&\longrightarrow &
\ P=(100;..\pm1..)\ etc.  \rightarrow 3[ ({\bf 
26},{\bf 3})_1+({\bf 1},{\bf 
3})_{-2}]\\
T&:& \ \
%&\longrightarrow & 
\ (P+V)=(1/3,1/3,-1/3..\pm 1..)\rightarrow 
27 ({\bf 26},{\bf 1})_{1/3}\\
T_{osc}&:&\ \
%& T(osc)&\longrightarrow &
(P+V)=(1/3,1/3,2/3,000..0) \rightarrow 3\times 27({\bf 1},\overline{\bf 
3})_{4/3}
\end{eqnarray}
Here the anomalous $U(1)_X$ is generated by the charge vector 
$Q_X=(1,1,1,0,..,0)$. Charges of fields are given by the
scalar products e.g.  $Q_X.(P+V)$.
In this case we have the sublattice:
\beq
\Lambda_0 \ =\ \left(  M(\pm 2/3) \ ,\ M\in {\bf Z}\right)  \ .
\eeq
The operators associated to the shortest charges in the lattice would be in 
this case
$[({\bf 26},{\bf 1})]^2_{2/3}$  and $[({\bf 1},{\bf 
3})({\bf 1},\overline{\bf 
3})]_{-2/3}$. 
There is a variety of choices which
can cancel the FI. One has $TrQ_X=24\times 36$ so one could cancel the FI with 
e.g.
$(1,{\bf 3})^3_{-2}$ or $([1,{\bf 3})({\bf 1},\overline{\bf 
3})]_{-2/3}$. Along these directions 
$SU(3)$ is also broken.
Note in this examples several scalar are required to cancel the FI, there is no 
singlet in the massless sector
to do the job.

\end{document}